\documentclass[11pt]{article}

\usepackage{amsmath}
\usepackage{amsfonts}
\usepackage{amssymb}
\usepackage{graphicx}
\usepackage{supertabular}
\usepackage{setspace}
\usepackage{xcolor}
\usepackage{array}
\usepackage{amsthm}
\usepackage{bm}
\usepackage{latexsym}
\usepackage{mathptmx}
\usepackage{hyperref}
\hypersetup{colorlinks=false}

\RequirePackage{cite}%
\renewcommand{\citeleft}{\bgroup\normalfont[}%
\renewcommand{\citeright}{]\egroup}%

\newcommand{\nin}{\noindent}

\newcommand{\be}{\begin{equation}}
\newcommand{\ee}{\end{equation}}
\newcommand{\ba}{\begin{eqnarray}}
\newcommand{\ea}{\end{eqnarray}}
\newcommand{\bal}{\begin{align}}
\newcommand{\eal}{\end{align}}

\newcommand{\dd}{{\rm d}}

\newcommand{\om}{\omega}
\newcommand{\al}{\alpha}
\newcommand{\la}{\lambda}
\newcommand{\bt}{\beta}
\newcommand{\ka}{\kappa}

\newcommand{\ga}{\gamma}
\newcommand{\ro}{\rho}

\newcommand{\ta}{\theta}

\newcommand{\Om}{\Omega}

\newcommand{\bw}{\begin{widetext}}
\newcommand{\ew}{\end{widetext}}

\def\rg{\rho_{\gamma}}
\def\rdm{\ro_{\text{DM}}}
\def\rf{\rho_{\phi}}
\def\of{\omega_{\phi}}

\def\P{3-fluid problem}
\def\p{3-fluid problem }

\begin{document}


\title{{\large\textbf{Phase-space analysis of the cosmological 3-fluid problem: Families of attractors and repellers}}}

\author{Mustapha Azreg-A\"{\i}nou
\\Ba\c{s}kent University, Department of Mathematics, Ba\u{g}l\i ca Campus, Ankara, Turkey}
\date{}

\maketitle

\begin{abstract}
We perform a phase-space analysis of the cosmological 3-fluid problem consisting of a barotropic fluid with an equation-of-state parameter $\gamma-1$, a pressureless dark matter fluid, plus a scalar field $\phi$ (representing dark energy) coupled to exponential potential $V=V_0\exp{(-\kappa\lambda\phi)}$. Besides the potential-kinetic-scaling solutions, which are not the unique late-time attractors whenever they exist for $\lambda^2\geq 3\ga$, we derive new attractors where both dark energy and dark matter coexist and the final density is shared in a way independent of the value of $\gamma >1$. The case of a pressureless barotropic fluid ($\gamma =1$) has a one-parameter family of attractors where all components coexist. New one-parameter families of matter-dark matter saddle points and kinetic-matter repellers exist. We investigate the stability of the ten critical points by linearization and/or Lyapunov's Theorems and a variant of the theorems formulated in this paper. A solution with two transient periods of acceleration and two transient periods of deceleration is derived.

\vspace{3mm}


\vspace{-3mm} \nin \line(1,0){430} 
\end{abstract}

\section{Introduction}

Observations have ever confirmed the so-called transient period of acceleration (TPA)~\cite{ob}, in this regard any realistic cosmological model should include component(s) with negative pressure. In this work we consider a spatially flat Friedmann-Roberston-Walker (FRW) cosmology where the content of the universe has two components with negative pressure and a pressureless component. This is the so-called \p consisting of a barotropic fluid with equation of state $p_{\ga}=(\ga-1)\rg$ where $0\leq\ga\leq2$, a pressureless dark matter (DM) density $\rdm$ and a dark-energy-scalar-field (DE) $\phi$ coupled to exponential potential $V(\phi)=V_0\exp{(-\ka\la\phi)}$ with equation of state $p_{\phi}=\of\rf$, $p_{\phi}=\dot{\phi}^2/2-V(\phi)$ and $\rf=\dot{\phi}^2/2+V(\phi)$. We assume that the three components are noninteracting. In this model, the barotropic fluid represents visible matter if $\ga\geq 1$ [radiation if $\ga=4/3$ or ordinary matter (baryons) if $\ga=1$]. For short, we will call this component matter ($\ga\geq 1$). In this cosmological 3-fluid model $-1\leq \of\leq 1$; however, from a physical point of view, there is no compelling reason to constrain the values of $\of$ to the interval [$-1,\,1$], in this regard it has been established that the teleparallel DE~\cite{para1} cosmological model allows for $\of<-1$~\cite{para2}.

Existing analytical methods~\cite{3fluid} have failed~\cite{And2} to produce exact solutions to the \P. Apart from some trivial solutions (power law inflationary solutions where the scale factor of the universe evolves as $a\propto t^m$  for $m>0$), no exact solution to the \p seems to exist to our knowledge.

In Ref.~\cite{num} we restricted ourselves to ordinary matter, where $1\leq\ga\leq2$, and to positive exponential potentials and resorted to numerical approach by which we derived new solutions to the \P. The solutions where classified hyperbolic and trigonometric according to the value of $\la$, this extends the classification made for the solutions to the cosmological 2-fluid problem~\cite{And} (consisting of a barotropic fluid plus a DE-scalar-field $\phi$ coupled to exponential potential), which were first derived in~\cite{Russo}. For the whole range of $\la$, we were able to construct solutions with one TPA where the universe undergoes the deceleration-TPA-eternal deceleration expansions. No solutions with two or more TPA's were found. However, as we shall see later, solutions with one TPA and a late-time eternal acceleration expansion do exist for the range $0\leq \ga < 2/3$ (which were not reported in Ref.~\cite{num}), where the universe undergoes the deceleration-TPA-TPD-eternal acceleration expansions (TPD: For transient period of deceleration). We shall also derive solutions with two TPA's and two TPD's for $\ga$ approaching from below 2/3. Solutions with many TPA's
and TPD's may exist too.

Phase-plane and -space analyses of autonomous differential equations lead to specific solutions that may provide the late-time attractors or the early-time repellers. Both type of solutions are interesting and provide rich insight into the evolution of the universe. Prior to the determination of the exact solutions to the cosmological 2-fluid problem~\cite{And}-~\cite{CTS}, phase-plane analyses of the 2-fluid problem were performed~\cite{p1}-~\cite{p4} (for a general procedure see~\cite{p3,psc}) and led to the discovery of potential-kinetic-scaling solutions, which are the unique late-time attractors whenever they exist for $\la^2>3\ga$.

We shall carry a phase-space analysis of the autonomous differential equations governing the dynamics of the \P. Among the conclusions we reach (1) the stability of the scalar field dominated solution for $\la^2\leq\min(3,3\ga)$, (2) the stability of the potential-kinetic-matter scaling solution for $\ga\leq 1$, which are quantitatively different from the corresponding results for the 2-fluid problem~\cite{p1}, and the existence of (3) new attractors (the potential-kinetic-DM scaling solution and the potential-kinetic-matter-DM scaling solution) and (4) new repellers and saddle points. In Sect.~\ref{sec2} we derive the autonomous differential equations of the \p and some other useful formulas. In Sect.~\ref{sec3} we discuss and extend the methods used for the stability analysis. In Sect.~\ref{sec4} we derive the critical points, investigate their stability and and their cosmologic implications, and construct numerically solutions with two TPA's and two TPD's as well as solutions with one TPA and a late-time eternal acceleration expansion. In Sect.~\ref{sec5} we discuss which physical scenarios are well fitted by the \P. We conclude in Sect.~\ref{sec6}.

\section{Autonomous differential equations of the \p \label{sec2}}

The three components being noninteracting each fluid satisfies a conservation equation of the form $T^{\mu\nu}_{\text{i}}{}_{;\nu}=0$ where $T^{\mu\nu}_{\text{i}}$ is the corresponding stress-energy tensor ($\text{i}=\ga,\,\phi,\,\text{DM}$). Keeping the relevant conservation equation for our analysis (corresponding to $\text{i}=\ga$) and using a similar notation as in~\cite{p1}, the dynamics of the three fluids in a spatially flat FRW universe, with a scale factor $a(t)$ and a Hubble parameter $H(t)=\dot{a}/a$, are governed by the very Eqs. (1) to (3) of~\cite{p1} upon slightly modifying the first equation by adding the contribution attributable to DM
\begin{align}
\label{1}& \dot{H}=-\frac{\ka^2}{2}(\rg+p_{\ga}+\rdm+\dot{\phi}^2)=-\frac{\ka^2}{2}(\ga\rg+\rdm+\dot{\phi}^2),\\
\label{2}& \dot{\rho}_{\ga}=-3H(\rg+p_{\ga})=-3H\ga\rg,\\
\label{3}& \ddot{\phi}=-3H\dot{\phi}-\frac{\dd V}{\dd \phi},
\end{align}
where $\dot{F}=\dd F/\dd t$. These equations are constrained by
\begin{equation}\label{4}
    H^2=(\ka^2/3)[\rg +\rdm +(\dot{\phi}^2/2)+V].
\end{equation}
From now on we consider only positive potentials $V(\phi)=V_0\exp{(-\ka\la\phi)}$ where $\la>0$. The dimensionless variables
\begin{equation}\label{5}
    x=\frac{\ka\dot{\phi}}{\sqrt{6}H}=\frac{\ka\phi'}{\sqrt{6}},\;y=\frac{\ka\sqrt{V}}{\sqrt{3}H},\;
    z=\frac{\ka\sqrt{\rg}}{\sqrt{3}H},\;w=\frac{\ka\sqrt{\rdm}}{\sqrt{3}H},
\end{equation}
where $\dot{F}=HF'$ and $F'=\dd F/\dd N$ ($N\equiv \ln a$), reduce the system~\eqref{1}-~\eqref{3} to the following system of three linearly independent autonomous differential equations
\begin{align}
\label{6}&x'=\sqrt{\frac{3}{2}} \lambda  y^2-3 x+\frac{3}{2} x [1+x^2-y^2+(\gamma -1) z^2],\\
\label{7}&y'=-\sqrt{\frac{3}{2}} \lambda  x y+\frac{3}{2} y [1+x^2-y^2+(\gamma -1) z^2],\\
\label{8}&z'=-\frac{3}{2} \gamma  z+\frac{3}{2} z [1+x^2-y^2+(\gamma -1) z^2],
\end{align}
where the variable $w$ is solved by
\begin{equation}\label{9}
    x^2+y^2+z^2+w^2=1,
\end{equation}
which follows from~\eqref{4}. It is worth mentioning that the expression in the square parentheses in the system~\eqref{6}-~\eqref{8} is positive or zero: $1+x^2-y^2+(\gamma -1) z^2=2x^2+\ga z^2+w^2$. To arrive at~\eqref{6}-~\eqref{8} we used
\begin{equation}\label{10a}
    H'/H=-3(2x^2+\ga z^2+w^2)/2.
\end{equation}

The equation governing the motion of $w$ is independent of $\la$
\begin{equation}\label{10}
    w'=3w [x^2-y^2+(\gamma -1) z^2]/2.
\end{equation}

In general (for all $\ga$), the constraint~\eqref{9} restricted the motion to within the unit solid 2-sphere of center at the origin: $x^2+y^2+z^2\leq 1$. A necessary formula for the stability analysis is readily derived upon combining~\eqref{6}, \eqref{7} and~\eqref{8} setting $x^2+y^2+z^2=r^2$
\begin{equation}\label{11}
    (r^2)'=3(r^2-1)(2x^2+\ga z^2-r^2).
\end{equation}

Another useful formula for the stability analysis and qualitative behavior of the solutions is derived upon eliminating the expression in the square parentheses in~\eqref{8} and~\eqref{10}
\begin{equation*}
    \frac{z'}{z}-\frac{w'}{w}=\frac{3}{2}(1-\ga)
\end{equation*}
leading to\footnote{Eq.~\eqref{12} is also derived upon combining Eqs. (18) and (19) of~\cite{3fluid}.}
\begin{equation}\label{12}
    z^2=L^2w^2a^{3(1-\ga)},
\end{equation}
where $L>0$ is a constant of integration. For a pressureless barotropic fluid ($\ga=1$), Eq.~\eqref{11} reduces to $z^2=L^2w^2$ which leads, using~\eqref{9} and setting $\ell =L/\sqrt{L^2+1}<1$, to
\begin{equation}\label{13}
    x^2+y^2+z^2/\ell^2=1.
\end{equation}
Thus for $\ga=1$, the motion happens on an ellipsoid of revolution around the $z$ axis in the phase space. The ellipsoid, which is inside the 2-sphere $x^2+y^2+z^2\leq 1$, does not contain all the trajectories for $\ga=1$; as we shall see, there are some equilibrium points inside and outside the ellipsoid; there are also other trajectories corresponding to $L=0$ ($\Rightarrow z=0$) and to $L=\infty$ ($\Rightarrow w=0$).

The relative densities are defined by $\Om_{\phi}\equiv x^2+y^2$, $\Om_{\ga}\equiv z^2$, $\Om_{\text{DM}}\equiv w^2$ and obey the conservation equation $\Om_{\phi}+\Om_{\ga}+\Om_{\text{DM}}=1$. Other relevant quantities are the parameter $\of=(x^2-y^2)/(x^2+y^2)$ which is constrained by $-1\leq \of \leq 1$ and the deceleration parameter $q\equiv -\ddot{a}/(aH^2)$ which is, by the field equations, the same as $-1-\dot{H}/H^2=-1-H'/H$ leading to
\begin{equation}\label{14}
    2q=1+3[x^2-y^2+(\gamma -1) z^2].
\end{equation}

Combining~\eqref{11} and~\eqref{14} we arrive at
\begin{align}
\label{14a}&\Om_{\text{DM}}'=(2q-1)\Om_{\text{DM}}\\
\label{14b}&\Om_{\ga}'=[(2q-1)-3(\ga -1)]\Om_{\ga}\\
\label{14c}&\Om_{\phi}'=(2q-1)(1-\Om_{\phi})+3(\ga -1)\Om_{\ga}\,.
\end{align}

\section{The critical points (CP's) -- Lyapunov's Stability and Instability Theorems (LST and LIT) \label{sec3}}

The CP's are the equilibrium points ($x_c,\,y_c,\,z_c,\,w_c$) in the phase space obtained upon solving the nonlinear algebraic equations $x'=0$, $y'=0$, $z'=0$, and $w'=0$. To determine the stability of the CP's we proceed to the linearization of the system~\eqref{6}-~\eqref{8} setting $x=X+x_c$, $y=Y+y_c$, $z=Z+z_c$ ($w=W+w_c$) where the new variables ($X,\,Y,\,Z$) still obey the full nonlinear system~\eqref{6}-~\eqref{8}. Upon linearization, the system~\eqref{6}-~\eqref{8} is brought to the matrix form:
\begin{equation}\label{15}
    (X',\,Y',\,Z')^{T}=J_c\cdot (X,\,Y,\,Z)^{T},
\end{equation}
where $(X,\,Y,\,Z)^{T}$ is the column matrix transpose of $(X,\,Y,\,Z)$ and $J_c$ is the $3\times3$ Jacobi matrix~\cite{b1}-~\cite{b3} evaluated at the CP ($x_c,\,y_c,\,z_c$):
\begin{equation}\label{16}
J_c=\begin{bmatrix}
J_{c\,11} & (-3 x_c+\sqrt{6} \lambda ) y_c & 3 (\gamma -1) x_c z_c \\
 (3 x_c-\sqrt{\frac{3}{2}} \lambda ) y_c & J_{c\,22} & 3 (\gamma -1) y_c z_c \\
 3 x_c z_c & -3 y_c z_c & J_{c\,33}
\end{bmatrix}
\end{equation}
where $2J_{c\,11}=3[-1+3 x_c^2-y_c^2+(\gamma -1) z_c^2]$, $2J_{c\,22}=3-\sqrt{6} \lambda  x_c+3 x_c^2-9 y_c^2+3 (\gamma -1) z_c^2$, $2J_{c\,33}=3[x_c^2-y_c^2+(\gamma -1) (3 z_c^2-1)]$.

The test for stability of almost linear systems~\cite{b1} states that~\cite{b1,b3} if (1) \textsl{all} the eigenvalues of the nonsingular matrix $J_c$ ($\det J_c\neq 0$) have negative real parts, then the CP is asymptotically stable but if (2) any eigenvalue has positive real part, then the CP is unstable. If some eigenvalues are zero ($\det J_c= 0$) or have zero real parts (and still $\det J_c\neq 0$), we will employ appropriate arguments (among which Lyapunov's Theorems, LST~\cite{b1}-~\cite{b3}, and LIT~\cite{b1}) for the determination of the stability of the corresponding CP as the above-mentioned test is no longer valid~\cite{b1,b3}. Mathematically speaking, we shall not make use of the notion of saddle points for a saddle CP is generically unstable. However, physically, we shall distinguish between a repeller and a saddle point.

LST assumes the existence of a continuously differentiable function $U(X,\,Y,\,Z)$ that is positive definite in a neighborhood $\mathcal{D}$ of the CP and has an \textsl{isolated} minimum at the CP, which is the origin in the new coordinates $(X,\,Y,\,Z)$: $(X_{\text{CP}},\,Y_{\text{CP}},\,Z_{\text{CP}})=(0,\,0,\,0)$. If further the derivative of $U$ along a solution curve, $U'=\partial_{i}U(X^i)'$ with $i=1\to3$ and $X^1=X,\,X^2=Y,\,X^3=Z$, is negative definite on $\mathcal{D}$ (except at the origin): $U'(X,\,Y,\,Z)<0$, then the CP is asymptotically stable. We are not concerned with the case where the CP is stable~\cite{b1}-~\cite{b3}.

LIT~\cite{b1} for 2-dimensional systems generalizes to higher dimensional systems in a straightforward way. It consists in finding a function $U(X,\,Y,\,Z)$ that is continues on a domain $\mathcal{D}$ containing the CP, which is assumed to be isolated. The Theorem assumes that $U(\text{CP})=0$ and that there is at least a point $P_0=(X_0,\,Y_0,\,Z_0)$ in each disc in $\mathcal{D}$, of center CP, where $U(P_0)>0$. If $U'$ is positive definite on $\mathcal{D}$ (except at the origin): $U'(X,\,Y,\,Z)>0$, then the CP is unstable.

The intuition behind LIT is as follows. Assume the above conditions are satisfied. In any disc $D_{\epsilon}$ in $\mathcal{D}$ select a solution curve that starts at\footnote{In a general problem, use $t$ instead of $N$.} $P_0$: $X(N=0)=X_0,\,Y(N=0)=Y_0,\,Z(N=0)=Z_0$. If the solution curve evolves from $P_0$ to, say, $P_1=(X_1,\,Y_1,\,Z_1)$ we must have $U(P_1)>U(P_0)>0$ since $U$ is increasing along the solution curve. Now, since $U(P_0)>U(\text{CP})=0$, this solution curve won't reach the CP in a finite or an infinite time $N$ (otherwise $U$ would decrease). Thus the CP is not asymptotically stable (an asymptotically stable point is a CP where any solution curve starting in its vicinity ends up at it as $N\to\infty$). Furthermore, the solution curve must leave the disc $D_{\epsilon}$ since, otherwise, as $N\to\infty$, $U\to\infty$ too, which is not possible as the continuity of $U$ on $D_{\epsilon}$ implies that it is bounded there. Thus, the CP is not stable; it must be unstable. LST works, in a sense, the other way around in that its hypotheses ensure that the solution curve approaches the CP as $N\to\infty$.

In LIT, $U$ need not be zero at the CP since one can add any positive or negative constant to $U$ without modifying the condition of stability, and $U(P_0)$ need not be positive\footnote{The LST and LIT were firstly formulated to deal with the stability of autonomous differential equations where the CP is the origin of the new coordinates $(X,\,Y,\,Z)$. This is no longer the case in other coordinate systems as $(x,\,y,\,z)$.}: It is sufficient to have $U(\text{CP})<U(P_0)$. A variant of LIT may be formulated as follows. If (1) $U(x,\,y,\,z)$ is continues on a domain $\mathcal{D}$ containing the CP, which is assumed to be isolated, (2) in every disc centered at the CP [here the CP is not necessarily the origin of the coordinates $(x,\,y,\,z)$], there exists some point $P_0=(x_0,\,y_0,\,z_0)$ such that $U(\text{CP})>U(P_0)$, (3) $U'$ is negative definite on $\mathcal{D}$, then the CP is unstable.

In cosmology both LST and LIT are very useful. One is interested to stable CP's or attractors where different solution curves end up at regardless of their initial conditions. One is also interested to unstable solutions or repellers which represent starting or intermediate events.

The main difficulty in applying LST and LIT is that there is no method how to find $U$. There are, however, some directions for that purpose~\cite{b3}. However, the construction of $U$ may be greatly simplified relying on the assumptions of LST and LIT concerning $\mathcal{D}$. This will be illustrated in the following discussion.

\section{The critical points (CP's) -- Stability analysis -- Cosmological implications \label{sec4}}

We have counted ten CP's labeled from $A$ to $J$. In the following we will provide the values of the CP's in the form $(x_c,\,y_c,\,z_c,\,w_c)$, determine their stability conditions and discuss their cosmological implications. The stability conditions are determined in terms of intervals of $\la$ and/or $\ga$ and are derived using the ``Hessian" test for stability as well as both LST and LIT. Particularly, the LST and LIT are employed to determine the stability conditions at the endpoints of the intervals of $\la$ and/or $\ga$, a task generally overlooked, skipped or difficult without use of the theorems~\cite{para2,p1,p2,psc,int3,int4}. We summarize our results in Table~\ref{Tab1}. As was mentioned earlier, no distinction is made in the text between a saddle point and an unstable CP; This distinction appears only in Table~\ref{Tab1}.

Following the classification made for the analytic solutions to the 2-fluid problem~\cite{And}, the solutions with $\la^2<6$ are called hyperbolic and those with $\la^2>6$ are called trigonometric. Due to different conventions, the value of $\la$ used in~\cite{num}, $\la_{\text{num}}$, is related to the value of $\la$ used in this work by $\la_{\text{num}}=\sqrt{3}\la$.

\begin{table}[h]
{\footnotesize
\begin{tabular}{|@{}c@{}|l|l|l@{}|l|l|l|}
  \hline
  \textbf{CP} & $\pmb{(x_c,\,y_c,\,z_c,\,w_c)}$ & \textbf{Existence} & \textbf{Stability} & $\pmb{\of}$ & $\pmb{2q}$ & $\pmb{\Om_{\ga}+\Om_{\text{DM}}}$ \\
  \hline
  \hline
  $A$ & $(0,\,0,\,0,\,1)$ & always & SP & $\nexists$ & 2 & 1 \\
  \hline
  $B_+$ & $(1,\,0,\,0,\,0)$ & always & Un & 1 & 4 & 0 \\
  \hline
  $B_-$ & $(-1,\,0,\,0,\,0)$ & always & Un & 1 & 4 & 0 \\
  \hline
   &  &  & $\la^2\leq\min(3,3\ga)$: AS &  &  &  \\
    $D$  & $(\la/\sqrt{6},\,\sqrt{1-(\la^2/6)},\,0,\,0)$ & $\la^2<6$ &  & $\frac{\la^2}{3}-1$ & $1+\la^2$ & 0 \\
      &  &  & $\min(3,3\ga)<\la^2<6$: SP &  &  &  \\
  \hline
   &  &  & $\ga=0$: AS &  &  &  \\
  $E$ & $(0,\,0,\,1,\,0)$ & always & $0<\ga <2$: SP & $\nexists$ & $3\ga-2$ & 1 \\
   &  &  & $\ga=2$: Un &  &  &  \\
  \hline
   & $(\cos\ta,\,0,\,\sin\ta,\,0)$ &  &  &  &  &  \\
  $F$ &  & $\ga=2$ & Un & 1 & 4 & $\sin^2\ta$ \\
      & ($0<\ta<\pi$) &  &  &  &  &   \\
  \hline
   & &  & $0\leq\ga\leq\frac{2}{9}$ \& $3\ga <\la^2$: SN &  &  &  \\
   & &  & $\frac{2}{9}<\ga\leq 1$ \& $3\ga <\la^2\leq\frac{24\ga^2}{9\ga-2}$: SN &  &  &  \\
  $G$ & $\Big(\sqrt{\frac{3}{2}}\frac{\ga}{\la},\,\sqrt{\frac{3}{2}}\frac{\sqrt{(2-\ga)\ga}}{\la},\,\frac{\sqrt{\la^2-3\ga}}{\la},\,0\Big)$ & $\la^2\geq 3\ga$ & $\frac{2}{9}<\ga\leq 1$ \& $\la^2>\frac{24\ga^2}{9\ga-2}$: SS & $\ga-1$ & $3\ga-2$ & $1-\frac{3\ga}{\la^2}$ \\
   & &  & $\ga\leq1$ \& $\la^2=3\ga\,$: AS &  &  &  \\
   & &  & $1<\ga<2$ \& $3\ga \leq\la^2$: SP &  &  &  \\
   & &  & $\ga=2$ \& $6 \leq\la^2$: Un &  &  &  \\
  \hline
   &  & & $\ga>1$ \& $3<\la^2\leq\frac{24}{7}$: SN &  &  &  \\
  $H$ & $\Big(\sqrt{\frac{3}{2}}\frac{1}{\la},\,\sqrt{\frac{3}{2}}\frac{1}{\la},\,0,\,\frac{\sqrt{\la^2-3}}{\la}\Big)$ & $\la^2\geq3$ & $\ga>1$ \& $\la^2>\frac{24}{7}$: SS & 0 & 1 & $1-\frac{3}{\la^2}$ \\
  &  & & $\ga=1$ \& $\la^2\geq3$: AS &  &  &  \\
  &  & & $\ga<1$ \& $\la^2\geq3$: SP &  &  &  \\
  \hline
   & $(0,\,0,\,\cos\ta,\,\sin\ta)$ &  &  &  &  &  \\
  $I$ &  & $\ga=1$ & SP & $\nexists$ & 1 & 1 \\
  & ($0<\ta<\pi/2$) &  &  &  &  &   \\
  \hline
   & & $\ga=1$ & $3<\la^2<\frac{24}{7}$: SN &  &  &  \\
  $J$ & $\Big(\sqrt{\frac{3}{2}}\frac{1}{\la},\,\sqrt{\frac{3}{2}}\frac{1}{\la},\,z_c,\,\sqrt{1-\frac{3}{\la^2}-z_c^2}\Big)$ & \& &  & 0 & 1 & $1-\frac{3}{\la^2}$ \\
   & & $\la^2>3$ & $\la^2>\frac{24}{7}$: SS &  &  &  \\
  \hline
\end{tabular}}
\caption{{\footnotesize Existence and stability of the critical points. \textsc{Nomenclature:} ``CP" for ``Critical Point", ``$\nexists$" for ``indefined", ``AS" for ``Asymptotically Stable", ``Un" for ``Unstable", ``SP" for ``Saddle Point", ``SN" for ``Stable Node", ``SS" for ``Stable Spiral".}}\label{Tab1}
\end{table}

\subparagraph{\pmb{$A=(0,\,0,\,0,\,1)$}.} For $\ga \neq 1$, $J_c$ has at least one positive eigenvalue: $\{-3/2, \,3/2,\,3(1-\ga)/2\}$. This CP is unstable. For $\ga = 1$, $J_c$ is singular. However, it is straightforward to show that in this case the CP is also unstable. This is achieved upon linearizating~\eqref{6} and~\eqref{7} in which case we obtain the eigenvalues $\mp 3/2$ of opposite signs.

\begin{figure}[h]
\centering
  \includegraphics[width=0.4\textwidth]{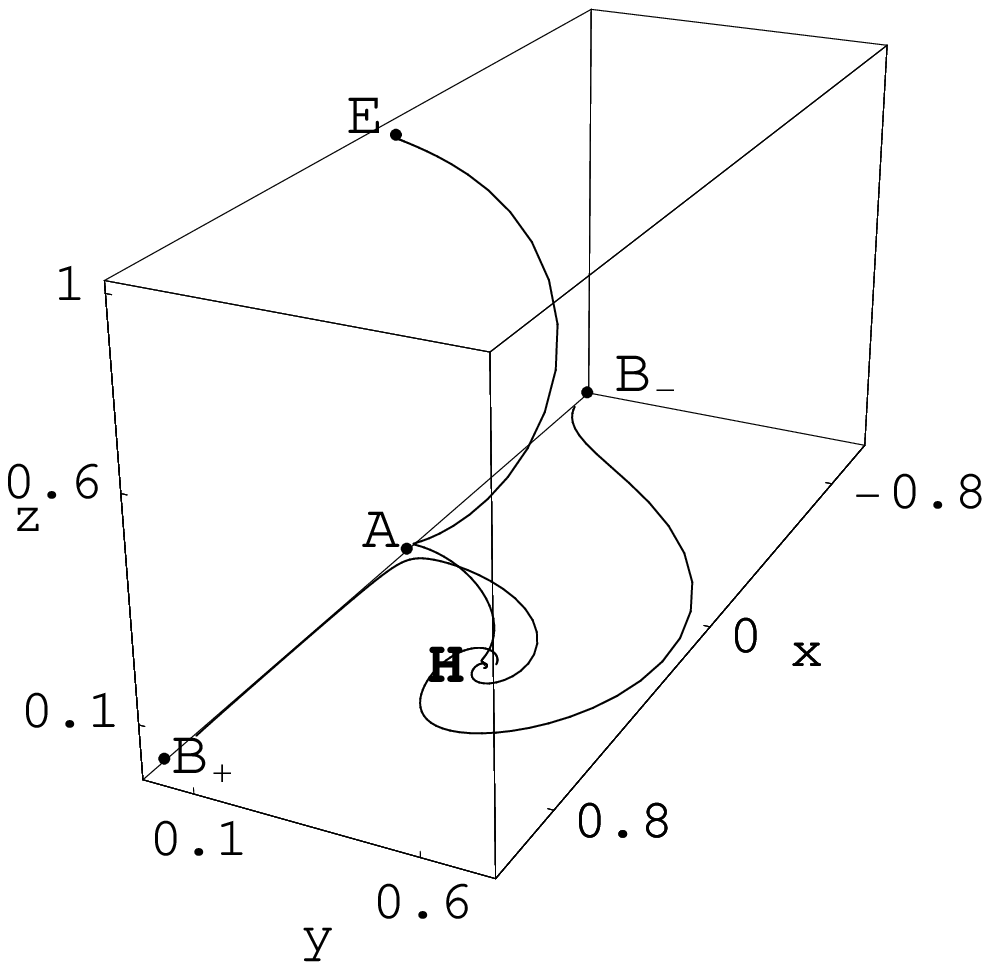} \includegraphics[width=0.4\textwidth]{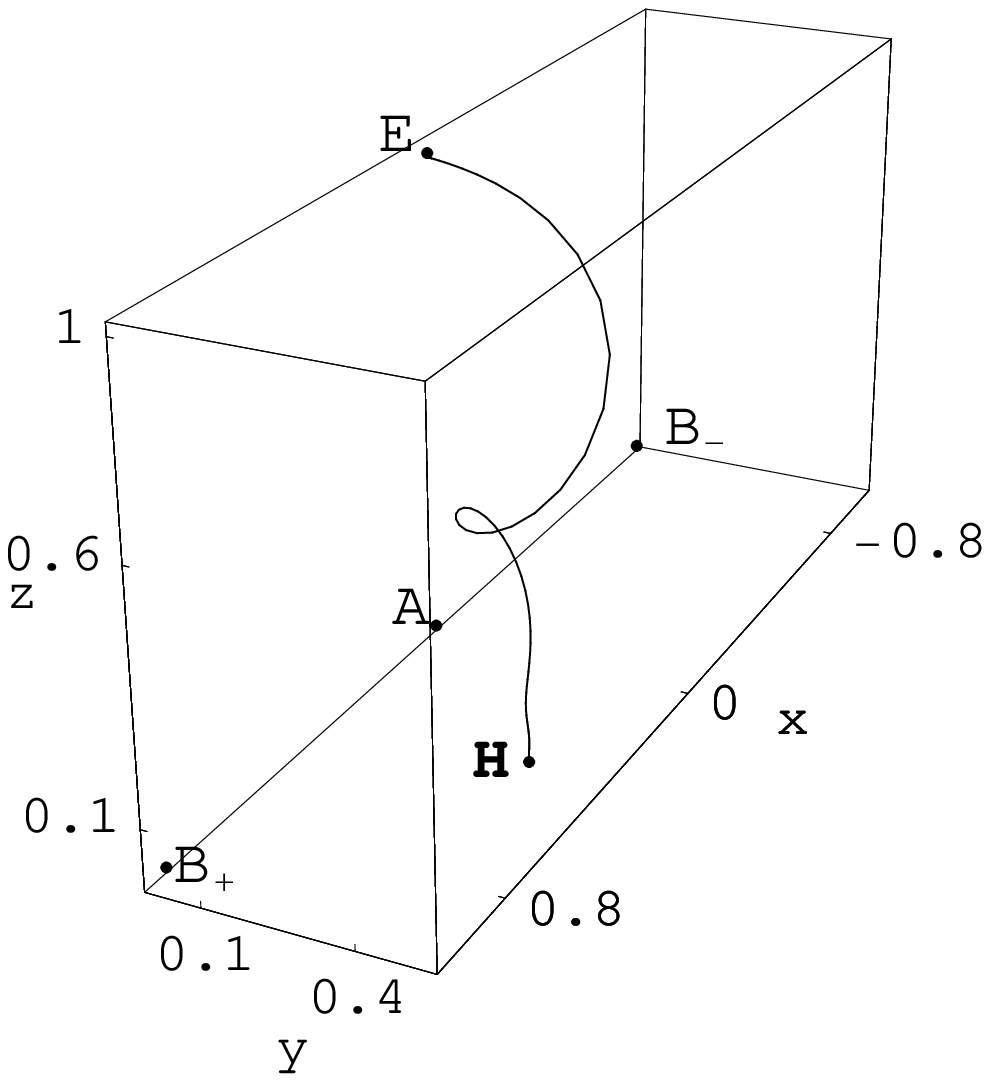}\\
  \caption{\footnotesize{(a): Left panel. Case $\la =3$, $\ga= 1.5$. For these values of the parameters, $H$ is the unique attractor. Solutions starting at $B_+$ and $E$ (for this value of $\ga$, $E$ is a saddle point) get very close to $A$, which is a saddle point. All solutions starting in the vicinity of $B_{\pm}$ and $E$ spiral to $H$. Those curves, which start in the vicinity of $B_{\pm}$ with $y\equiv 0$ and $z\equiv 0$ or in the vicinity of $E$ with $x\equiv 0$ and $y\equiv 0$, end up at $A$. Since $A$ is unstable, any perturbations in the values of the coordinates cause the solution curve to continue its journey to $H$. (b): Rigt panel. Case $\la =3$, $\ga= 4/3$.}}\label{Fig1}
\end{figure}

Cosmologically, the only solution curves that may reach $A$ emanate from $B_{\pm}$ with $y\equiv 0$ and $z\equiv 0$; otherwise, some solution curves (only those emanating from $B_+$) may just get close to it, but do not cross it, as shown in Fig.~\ref{Fig1}. At this CP, all densities vanish for a dominant DM component $\Om_{\text{DM}}=1$, $\of$ is an indeterminate, and the universe undergoes a decelerated expansion with $q=1/2$.

\subparagraph{\pmb{$B_+=(+1,\,0,\,0,\,0)$}, \pmb{$B_-=(-1,\,0,\,0,\,0)$}.} The matrix $J_c$ has the eigenvalues $\{3,\,(6-\epsilon\sqrt{6}\la)/2,\,3(2-\ga)/2\}$ where $\epsilon=1,\,-1$ for $B_+,\,B_-$, respectively, so they are generically unstable. They are also unstable in the special case $\ga =2$ where $J_c$ is singular\footnote{For $B_+$, if $\ga =2$ and $\la =\sqrt{6}$, we conclude to the instability upon applying LIT with $U=aX^2$ and $a>0$. The instability of the case $\ga =2$ and any $\la$ can also be achieved considering~\eqref{11} which becomes $(r^2)'=3(r^2-1)(r^2-2y^2)$. A solution curve that starts near the CP has $r<1$ (the CP is on the sphere $r=1$ and all solution curves are inside the sphere). Since $y^2=Y^2\ll r^2\approx 1$, we have $(r^2)'<0$ and thus the solution curve moves in the direction of decreasing $r$ and never returns back to the CP where $r=1$, which is then unstable. This is a first application of a variant of LIT which was formulated in the previous section.\label{var}} as can be concluded from the linearization of~\eqref{6} and~\eqref{7}.

\begin{figure}[h]
\centering
  \includegraphics[width=0.5\textwidth]{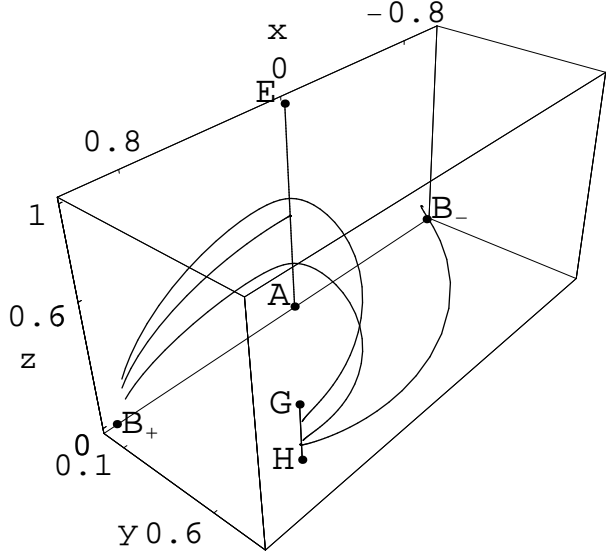}\\
  \caption{\footnotesize{Case $\la =1.8$, $\ga= 1$. For these values of the parameters, the vertical line $J$, through the end-points $H=(\sqrt{3/2}/\la,\,\sqrt{3/2}/\la,\,0)$ and $G=(\sqrt{3/2}/\la,\,\sqrt{3/2}/\la,\,\sqrt{1-(3/\la^2)})$, is the unique family of attractors ($H$ and $G$ are locally stable for $\la =1.8$, $\ga= 1$). $I$ is the line through $A=(0,\,0,\,0)$ and $E=(0,\,0,\,1)$. There is a curve starting in the vicinity of $B_-$ which converges to a point on the line $J$. There are three curves starting in the vicinity of $B_+$. The upper and lower curves approach the line $I$ then converge to different points on the line $J$. The intermediate curve, which corresponds to $y\equiv 0$ converges to the line $I$, which is a set of saddle points; any perturbation in the value of $y$ causes this curve to end up at any point on the line $J$. [In this caption the coordinates of the CP's have been given on the form $(x_c,\,y_c,\,z_c)$.]}}\label{Fig2}
\end{figure}

These are the repellers, as shown in Fig.~\ref{Fig1} and Fig.~\ref{Fig2}, with a dominant kinetic energy, a decelerated expansion $q=2$, and $\of=1$. For a steep potential, $\la>\sqrt{6}$, $B_+$ is a saddle point.

\subparagraph{\pmb{$D=(\la/\sqrt{6},\,\sqrt{1-(\la^2/6)},\,0,\,0)$}.} This CP exits for $\la^2<6$ (the case $\la^2=6$ leads to the previous case). The eigenvalues of $J_c$ are: $\{(\la^2-3\ga)/2,\,\la^2-3,\,(\la^2-6)/2\}$. In the case $\det J_c\neq 0$, the CP is asymptotically stable for $\la^2<\min(3,3\ga)$ and unstable for $\min(3,3\ga)<\la^2<6$ [this includes the cases ($\la^2=3$ and $\ga<1$) and ($\la^2=3\ga$ and $\ga>1$)]. If $\det J_c=0$, it is asymptotically stable in the cases ($\la^2=3$ and $\ga>1$) and ($\la^2=3\ga$ and $\ga<1$) upon linearizing~[\eqref{7} and~\eqref{8}] and [\eqref{6} and~\eqref{7}], respectively.

There remains the case $\la^2=3$ and $\ga=1$ where we expect the CP [in this case $D=(1/\sqrt{2},\,1/\sqrt{2},\,0,\,0)$] to be asymptotically stable. We apply LST and select $U$ of the form: $U=a(X+Y)^2+(b-a)Y^2+cZ^2$, which is positive definite if $0<a<b$ and $0<c$. The CP is an isolated minimum of $U$ with $U(CP)=0$. The directional derivative along the solution curves, $U'=\partial_{i}U(X^i)'$ with $i=1\to3$ and $X^1=X,\,X^2=Y,\,X^3=Z$, is evaluated using the r.h.s's of~\eqref{6}, \eqref{7} and~\eqref{8} after converting to new coordinates ($X,\,Y,\,Z$):
\begin{equation}
    U'=F(X,Y,Z) \; \text{ with }\; F=-3(b-a)Y^2+O[(X^i)^3],
\end{equation}
which is negative definite in the vicinity of the CP. We choose $\mathcal{D}$ to be any neighborhood of the CP (including the CP), where $U'<0$, and not including other points of the surface $S:\, F(X,Y,Z)=0$. This way we make $U$ negative definite\footnote{This evokes the pendulum problem~\cite{b2}. Even if $\mathcal{D}$ were to include other points on the surface $S$ (in which case $U'$ would be negative semidefinite), we would still conclude that the CP is asymptotically stable (and not just stable).} in $\mathcal{D}$. With these choices we satisfy the hypotheses of LST, so the CP with the case $\la^2=3$ and $\ga=1$ is asymptotically stable.

For $\la^2\leq\min(3,3\ga)$, this CP is an attractor with a dominant scalar field component $\Om_{\phi}=1$, $\of=-1+\la^2/3\leq 0$ and a decelerated expansion $2q=1+\la^2$.

\subparagraph{\pmb{$E=(0,\,0,\,1,\,0)$}.} From the set of the eigenvalues of $J_c$, $\{-3(2-\ga)/2,\,3\ga/2,\,3(\ga-1)\}$, the CP is generically unstable if $\det J_c\neq 0$.

Now, consider the case where $J_c$ is singular ($\det J_c=0$). In the special case $\ga =2$, the CP is unstable for the linearization of~\eqref{7} and~\eqref{8} leads to $Y'=3Y$, $Z'=3Z$. The same conclusion is achieved from $(X^2+Y^2+Z^2)'=3(Y^2+Z^2)+\cdots >0$. For $\ga =1$ the CP is also unstable by LIT or upon linearizing~\eqref{6} and~\eqref{7} which results in the eigenvalues $\pm 3/2$ of opposite signs. The case $\ga=0$ is stable since we have $(X^2+Y^2+Z^2)'=-3(Y^2+Z^2)+\cdots <0$.

\begin{figure}[h]
\centering
  \includegraphics[width=0.7\textwidth]{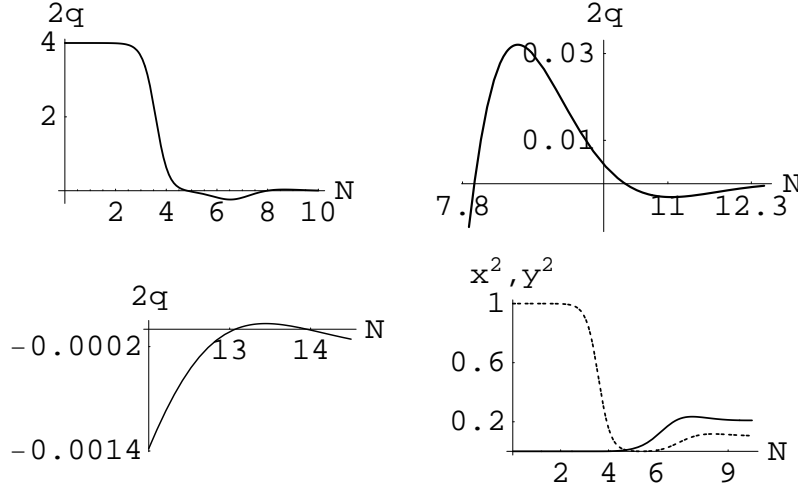}\\
  \caption{\footnotesize{Case $\la =\sqrt{6.3}$, $\ga= 0.6666$. For our initial conditions, at $N=0$, we took $x_0=-0.9999997$, $y_0=4.3569\times 10^{-12}$, $z_0=\sqrt{1-x_0^2-y_0^2}$. (Upper and lower left plots) Twice the deceleration parameter $2q$. (Lower right plot) The kinetic and potential relative densities, $x^2$ (dashed line) and $y^2$ (continuous line). The parameter $2q$ crosses the $N$ axis at: $N_1=4.82333$, $N_2=7.984$, $N_3=10.3342$, $N_4=13.0676$ and $N_5=13.9537$. This solution has thus two TPA's and two TPD's: The first and second TPA's are observed in the intervals $N_1<N<N_2$ and $N_3<N<N_4$, respectively, and the first and second TPD's are observed in the intervals $N_2<N<N_3$ and $N_4<N<N_5$, respectively. The first TPA starts at $N=N_1$, which is the moment where $x^2\simeq 0$ and $y^2\simeq 0$, that is the corresponding point on the solution curve is near $E$.}}\label{Fig3}
\end{figure}

\begin{figure}[h]
\centering
  \includegraphics[width=0.7\textwidth]{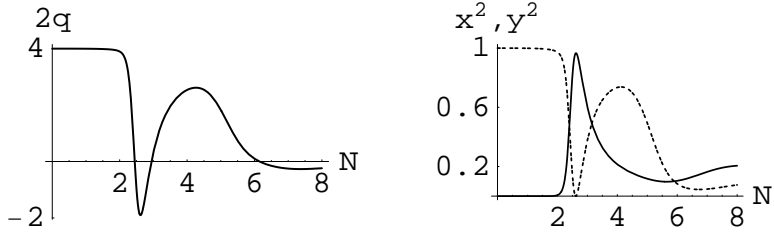}\\
  \caption{\footnotesize{Case $\la =\sqrt{6.3}$, $\ga= 0.6$. For our initial conditions, at $N=0$, we took $x_0=-0.999975$, $y_0=4.3569\times 10^{-7}$, $z_0=0.000479$. (Left plot) Twice the deceleration parameter $2q$. (Right plot) The kinetic and potential relative densities, $x^2$ (dashed line) and $y^2$ (continuous line). The parameter $2q$ crosses the $N$ axis at: $N_1=2.44678$, $N_2=2.94285$ and $N_3=6.14727$. This solution has thus one TPA and one TPD: The TPA is observed in the interval $N_1<N<N_2$ and the TPD is observed in the interval $N_2<N<N_3$. The TPA starts at $N_1$, which is the moment where $x^2$ is minimum and $y^2$ is maximum.}}\label{Fig4}
\end{figure}

Thus, $E$ is a matter dominant attractor ($\Om_{\ga}=1$) if $\ga =0$, a saddle point if $0<\ga <2$, and a repeller if
$\ga=2$. With the parameter $\of$ remains undetermined, the state of the universe at $E$ undergoes a decelerated expansion if $2/3<\ga \leq 2$ or an accelerated expansion if $0\leq \ga < 2/3$. This is a novel point because one may have a TPA without necessary having (at the same time) a minimum kinetic energy and a maximum potential energy as in the case of the 2-fluid  problem~\cite{Russo,num}. In fact, at $E$ both kinetic and potential energies are zero.
Thus if, for $0\leq \ga < 2/3$, a solution curve approaches the saddle point $E$ then deviates to a CP (this would be the CP $G$), the TPA there (at $E$) may last longer than the TPA occurring away from saddle points. This is because a saddle point behaves partly as an attractor and partly as a repeller. This is in fact the case in Fig.~\ref{Fig3} that is a plot, for $\la =\sqrt{6.3}$ and $\ga= 0.6666$, of twice the deceleration parameter, $2q$, and the kinetic and potential relative densities, $x^2$ (dashed line) and $y^2$ (continuous line). The parameter $q$ crosses the $N$ axis at: $N_1=4.82333$, $N_2=7.984$, $N_3=10.3342$, $N_4=13.0676$ and $N_5=13.9537$. This solution has thus two TPA's and two TPD's: The first and second TPA's are observed in the intervals $N_1<N<N_2$ and $N_3<N<N_4$, respectively, and the first and second TPD's are observed in the intervals $N_2<N<N_3$ and $N_4<N<N_5$, respectively. The first TPA starts at $N=N_1$, which is the moment where $x^2\simeq 0$ and $y^2\simeq 0$ [$x(N_1)=-0.075$, $y(N_1)=0.106$, $z(N_1)=0.991$], that is the corresponding point on the solution curve is near $E$. The graph of $2q$ continues to oscillate for $N>N_5$ below the line $q=0$, this is a sign that solutions with many TPA's and TPD's may exist if careful choice of the parameters is carried out. Fig.~\ref{Fig4} is a similar plot with different inputs $\la =\sqrt{6.3}$ and $\ga= 0.6$. It is a solution with one TPA and one TPD which depicts a case with a TPA starting at the moment where $x^2$ is minimum and $y^2$ is maximum.
\begin{figure}[h]
\centering
  \includegraphics[width=0.5\textwidth]{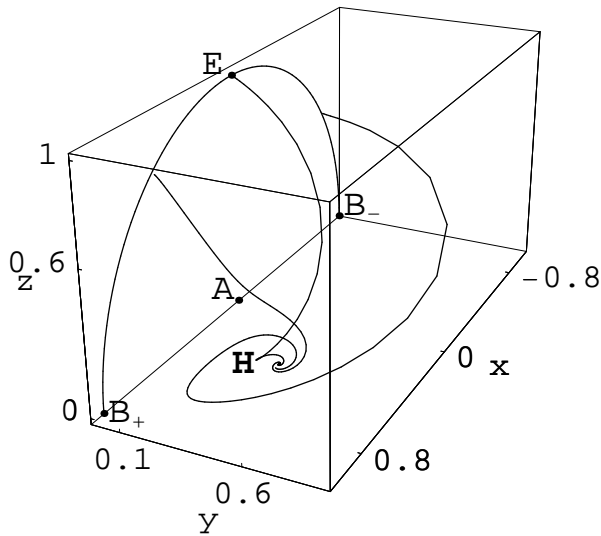}\\
  \caption{\footnotesize{Case $\la =3$, $\ga= 2$. For these values of the parameters, $H$ is the unique attractor. The circle through $B_+$, $E$ and $B_-$ is the one-parameter family $F$ of repellers plus $B_{\pm}$. Any curve starting in the vicinity of this kinetic-matter repeller ends up at $H$.}}\label{Fig5}
\end{figure}
\begin{figure}[h]
\centering
  \includegraphics[width=0.5\textwidth]{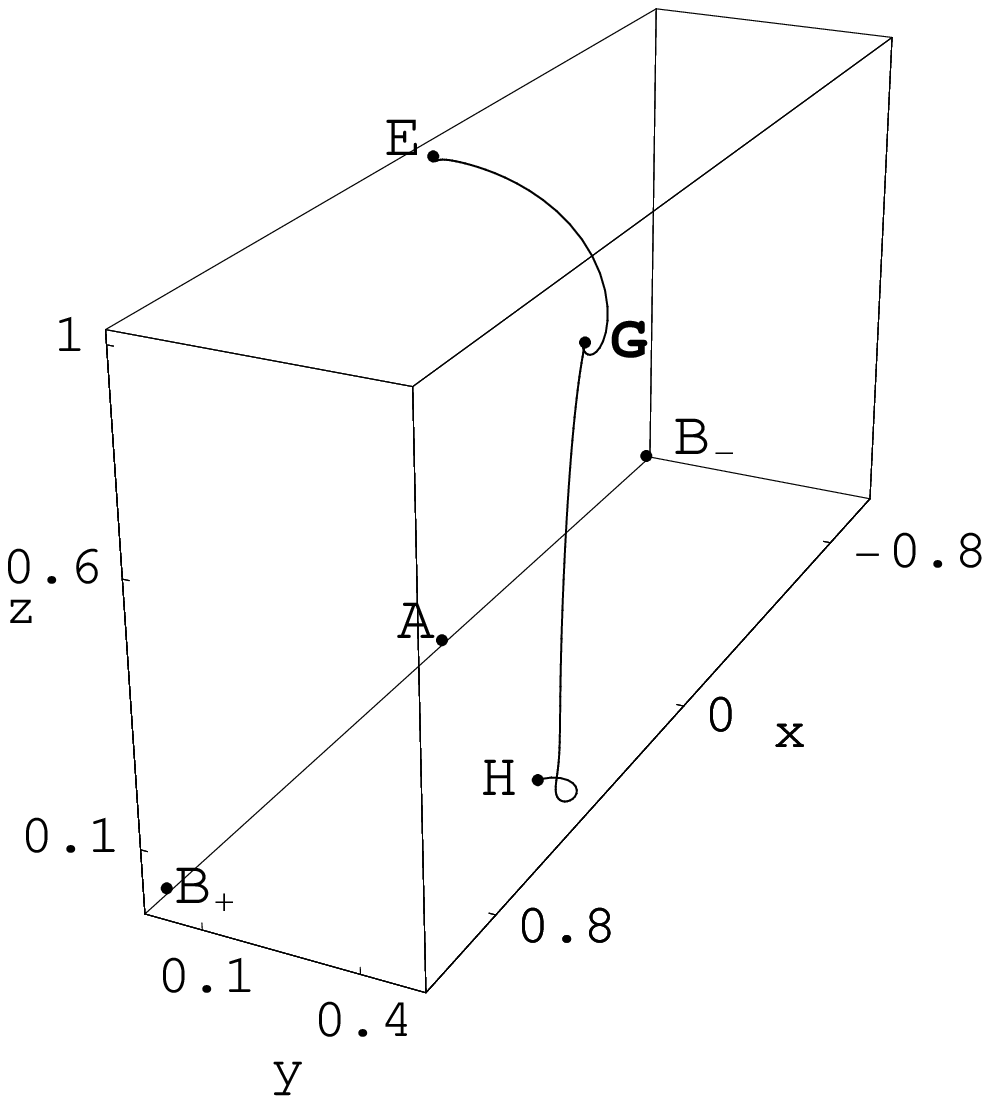}\\
  \caption{\footnotesize{Case $\la =\sqrt{6.3}$, $\ga= 0.6666$. Two solution curves starting from $E$ and $H$ and ending up at $G$.}}\label{Fig6}
\end{figure}

\subparagraph{\pmb{$F=(\cos\ta,\,0,\,\sin\ta,\,0)$}, \pmb{$0<\ta<\pi$}, \pmb{$\ga =2$}.} This unstable CP generalizes $B_{\pm}$ in that $\Om_{\ga}=\sin^2\ta$ may assume any value between 0 and 1; it also generalizes $E$.

As $J_c$ is singular, it is not possible to draw any conclusion concerning stability by linearization of the system~\eqref{6}-~\eqref{8} for this CP. With $\ga =2$, Eq.~\eqref{11} becomes $(r^2)'=3(r^2-1)(r^2-2y^2)$. A solution curve that starts near the CP has $r<1$ (the CP is on the sphere $r=1$ and all solution curves are inside the sphere). Since $y^2=Y^2\ll r^2\approx 1$, we have $(r^2)'<0$ and thus the solution curve moves in the direction of decreasing $r$ and never returns back to the CP where $r=1$, which is then unstable. This gives another application of a variant of LIT (see footnote~\ref{var}).

Since $\ta$ is not constrained, $F$ is a new one-parameter family of kinetic-matter repellers. With a stiff equation of state $\ga =2$, the initial density is shared between the barotropic fluid and DE, $\of =1$, and $q=2$ (decelerated expansion). Solution curves starting from $F$, which is represented by a semicircle in Fig.~\ref{Fig5}, reach $H$.

\subparagraph{\pmb{$G=(\sqrt{3/2}(\ga/\la),\,\sqrt{3/2}\sqrt{(2-\ga)\ga}/\la,\,\sqrt{\la^2-3\ga}/\la,\,0)$}.} The corresponding solution for the 2-fluid problem~\cite{p1} is a potential-kinetic scaling solution the stability of which does not depend on the value of $\ga$. For the \p we rather have a potential-kinetic-matter scaling solution, its stability depends on $\ga$ as we shall see later soon.

The eigenvalues depend on both ($\la,\,\ga$): $(3/4)\{4(\ga-1),\,\ga-2-\al,\,\ga-2+\al\}$ where we define $\al\equiv \sqrt{(2-\ga)[24\ga^2+\la^2(2-9\ga)]}/\la$. This CP exists for $\la^2\geq 3\ga$ only. If $\det J_c\neq 0$, it is asymptotically stable for $\ga<1$ and $\la^2> 3\ga$. Furthermore, this CP is (a) a stable node for $0\leq\ga\leq 2/9$ for all\footnote{This subcase exists also for the the 2-fluid problem but was not derived in~\cite{p1,para2}. In this subcase ($0\leq\ga\leq 2/9$), $\la^2$ need not be smaller than $24\ga^2/(9\ga-2)$.} $\la^2> 3\ga$ (the case $\ga=0$ leads to $E$ discussed above), (b) a stable node for $2/9<\ga<1$ and $3\ga<\la^2\leq 24\ga^2/(9\ga-2)$, and (c) a stable spiral if $2/9<\ga<1$ and $\la^2> 24\ga^2/(9\ga-2)$. The CP is unstable for $1<\ga<2$ and $\la^2> 3\ga$.

If $\det J_c=0$, this CP is asymptotically stable for $\ga=1$ and $\la^2> 3$ since the linearization of~\eqref{6} and~\eqref{7} provides two negative eigenvalues: $-(3/4)(1\pm \bt)$ with $\bt\equiv \sqrt{24-7\la^2}/\la$ (a stable node for $3<\la^2\leq 24/7$ and a stable spiral for $\la^2>24/7$). For the remaining cases where $J_c$ is singular ($\ga = 2$ or $\la^2= 3\ga$), the CP is unstable for $\ga=2$ and $\la^2\geq 6$ since near it we establish: $(X^2+Y^2+Z^2)'=6(\sqrt{\la^2-6}Z+\sqrt{6}X)^2/\la^2+\cdots >0$. For $\la^2= 3\ga$, the eigenvalues which result from the linearization of~\eqref{6} and~\eqref{7} are proportional to $\ga-2$ and $\ga-1$, ensuring asymptotic stability for $\ga<1$ and instability for $1<\ga<2$. For the case $\ga=1$ and $\la^2=3$, where $\det J_c=0$, we recover the CP $D=(1/\sqrt{2},\,1/\sqrt{2},\,0,\,0)$ which has been shown to be asymptotically stable.

The fact that $\ga\leq 1$ ---to ensure asymptotic stability of the CP--- results in $\of=\ga -1\leq 0$, while in the case of the 2-fluid problem $\of$, still given by the same formula, may have both signs. $\Om_{\phi}$ and $\Om_{\ga}$ depend on both ($\la,\,\ga$): $\Om_{\phi}=3\ga/\la^2$, $\Om_{\ga}=1-\Om_{\phi}$. With $2q=1+3(\ga-1)$, the state of the universe approaching this stable point may undergo a decelerated expansion if $2/3<\ga \leq1$ or an accelerated expansion if $0\leq \ga < 2/3$. $G$ is a saddle point for $1<\ga<2$.

\subparagraph{\pmb{$H=(\sqrt{3/2}/\la,\,\sqrt{3/2}/\la,\,0,\,\sqrt{\la^2-3}/\la)$}.} Here again the eigenvalues depend on both parameters ($\la,\,\ga$): $(3/4)\{6(1-\ga),\,-1-\bt,\,-1+\bt\}$. This CP exists for $\la^2\geq 3$ only and it is asymptotically stable for $\ga>1$ and $\la^2>3$. The CP is (a) a stable node if $3<\la^2\leq24/7$ (with $\ga>1$) or (b) a stable spiral if $\la^2>24/7$ (with $\ga>1$). If $\det J_c=0$, it is asymptotically stable for $\ga=1$ and $\la^2>3$ as the linearization of~\eqref{6} and~\eqref{7} leads to the same eigenvalues $(3/4)\{-1-\bt,\,-1+\bt\}$. For the case $\ga=1$ and $\la^2=3$ we recover the CP $D=(1/\sqrt{2},\,1/\sqrt{2},\,0,\,0)$ which has been shown to be asymptotically stable.

The relevant parameters are $\Om_{\phi}=3/\la^2$ which depends only on $\la$, $\of=0$, $\Om_{\text{DM}}=1-\Om_{\phi}$, and $q=1/2$.

For the 2-fluid problem, $G$ is the unique attractor for $\la^2\geq 3\ga$ (for all $\ga$)~\cite{p1}. But $G$ depends on $\ga$, this means that the end-behavior of the solution depends on the nature of the barotropic fluid. We have seen that, for the \P, $G$ is no longer stable for $1<\ga\leq 2$ but $H$ is, which is a new attractor and does not depend on $\ga$. Thus, no matter the barotropic fluid equation of state is, the universe's evolution ends up at the same state provided $1<\ga\leq 2$. As we shall se below, for $\ga=1$ ($\la^2\geq 3\ga$), there is a line (a one-parameter family) of attractors all represented by the CP $J$.

\subparagraph{\pmb{$I=(0,\,0,\,\cos\ta,\,\sin\ta)$}, \pmb{$0<\ta<\pi/2$}, \pmb{$\ga =1$}.} $J_c$ is singular, however, the eigenvalues which result from the linearization of~\eqref{6} and~\eqref{7} are $\pm 3/2$, ensuring instability.

This unstable CP generalizes $E$ and $A$ in that $\Om_{\ga}$ may assume any value constrained by $\Om_{\ga}+\Om_{\text{DM}}=1$, with $q=1/2$ and $\of$ remains undetermined. Since $\ta$ is not constrained, $I$ is a new one-parameter family of saddle points where only matter and DM are the nonvanishing components. $I$ is represented by a vertical line in the phase diagram, which is the line $AE$ of Fig.~\ref{Fig2}.

\subparagraph{\pmb{$J=(\sqrt{3/2}/\la,\,\sqrt{3/2}/\la,\,z_c,\,\sqrt{1-(3/\la^2)-z_c^2})$}, \pmb{$\ga =1$}.} The CP exists only for $\la^2>3$. With $\det J_c=0$, the CP is however asymptotically stable since the linearization of~\eqref{6} and~\eqref{7} provides the two negative eigenvalues: $-(3/4)(1\pm \bt)$ (a stable node for $3<\la^2\leq 24/7$ and a stable spiral for $\la^2>24/7$).

Since $\ta$ is a free parameter, $J$ is a new one-parameter family of attractors in which all components coexist: It is a potential-kinetic-matter-DM scaling solution where $\of=0$, $q=1/2$, $\Om_{\phi}=3/\la^2$, and $\Om_{\ga}+\Om_{\text{DM}}=1-(3/\la^2)$. $\Om_{\ga}$ and $\Om_{\text{DM}}$ are both arbitrary and smaller than $1-(3/\la^2)$. According to the last and most accurate observations~\cite{obs}, $\Om_{0\,\phi}=0.721\pm 0.015$ (corresponding to $\la^2 =4.161$), thus $\Om_{0\,\ga}\leq 0.279\pm 0.015$.

In Fig.~\ref{Fig2}, $I$ is any point on the line through $A$ and $E$. Only solution curves emanating from $B_{\pm}$ and $I$ (including $A$ and $E$) may reach $J$, which is represented by a vertical line in the phase diagram, this is the line $HG$ of Fig.~\ref{Fig2}.

\section{Fitting the 3-fluid model \label{sec5}}

As stated in the Introduction, any realistic model should include at least one component with a negative pressure to account for a TPA~\cite{ob}. For that purpose, many theoretical models have been suggested the simplest of which is the so-called $\Lambda$CDM where the component with negative pressure is a vacuum energy (the cosmological constant). This model results in a constant DE equation of state, $\om_{\text{DE}}=-1$, and the coincidence problem. The next generation of models, which consider two noninteracting fluids~\cite{And,Russo,p1,p2,2mod}, have introduced a scalar field (quintessence) to generalize the $\Lambda$CDM model. They have emerged to tackle the coincidence problem and to provide a variable DE equation of state. Models where ordinary matter and DE interact have also emerged~\cite{psc,int1}.

While there is no observational evidence of the existence of any interaction between the two dark components, some authors, however, arguing that the amounts of DE and DM are comparable at the present age of the universe, have anticipated that and formulated 2- and 3-fluid problems with DE-DM~\cite{int3,int4,int2} or DE-matter-radiation~\cite{psc} interaction terms. Of course, these models reduce to the \p with no interaction terms if appropriate constraints are further imposed. In Ref.~\cite{int4}, the authors considered a DE-DM interaction with baryons uncoupled and radiation redshifted away or neglected. Thus, their model applies to the epoch beyond the matter-radiation decoupling, which corresponds to a redshift $\text{z}_{\text{dec}}=1099.9$~\cite{books}, and it reduces to the \p upon setting the DE-DM interaction coupling constant $\bt=0$~\cite{int4} (with this constraint, the model corresponds to ours with $\ga=1$, as we shall see below in this section). In contrast, the authors of Ref.~\cite{psc}, considering again a flat FRW, included radiation in, but dropped baryons from, their DE-DM-radiation model to allow for a deeper investigation of the universe's dynamics during the radiation dominant era. They considered a non-minimally and non-constant coupling, inspired from Scalar Tensor Theories (STT), the value of which depends on the trace of the energy-momentum tensor of the background component. Since radiation is traceless, it remains decoupled from the DE-DM system they investigated. The authors presented a general procedure for dynamical analysis of the STT inspired DE-DM interactions. Their model reduces to the \p if their DE-DM interaction coupling function~\cite{psc} $\chi(\phi)=\text{constant}$ and their DM parameter $\ga_{\text{~\cite{psc}}}=1$ (with these constraints, the model corresponds to ours with $\ga=4/3$, as we shall see below in this section).

In most of the above-mentioned models, the potential functions associated with DE and/or the interaction terms have been derived or chosen, relying partly on some physical assumptions, so that the problems remain analytically tractable (even though no nontrivial exact analytic solutions have been found so far), among which we find the \p we are considering here with no interaction terms. However, by neglecting all types of interactions, particularly that of visible matter, we restrict the application of the 3-fluid model to beyond the epoch of matter-radiation decoupling ($\text{z}_{\text{dec}}=1099.9$~\cite{books}). Thus, for $\text{z}<\text{z}_{\text{dec}}$, the model fits well the following three physical scenarios based solely on the value of $\ga$.

\begin{enumerate}
  \item $\ga=4/3$. In this case the components of the universe are regrouped in a way that the barotropic fluid represents radiation, the DM and baryons together make up the pressureless component with a total relative density $\Om_{0}=0.279$, a baryonic density $\Om_{0\,b}=0.04-0.05$ and a DE density $\Om_{0\,\phi}=0.721$~\cite{obs} at the present age.

      The only stable attractor corresponding to this application is $H$ provided $\la^2>3$. The application has three saddle points $G$, provided $\la^2\geq 4$, $E$, and $A$. In this case, the model describes the evolution of the universe starting from $E$ (or from $G$ if $\la$ is large), where radiation is dominant, passing through or approaching $A$, where the pressureless component (matter) is dominant, and ending up at $H$, where the universe content is shared between DE and DM (which becomes dominant for large $\la$), as Fig~\ref{Fig1} depicts.
  \item $\ga=1$. The epoch of matter-radiation equality~\cite{books} corresponds to a redshift $\text{z}_{\text{eq}}=24000\Om_0h^2-1=3470.2$ where we take $h=0.72$. With $\text{z}<\text{z}_{\text{dec}}$ it is a good approximation to neglect radiation. It is now easy to see that the pressureless barotropic fluid represents baryons. In fact, at late times (but well before formation of structures), as the temperature drops by the effect of the expansion, baryons behave as a nonrelativistic ``monoatomic" ideal gas with pressure $p_b=n_bk_{\text{B}}T_b$ and mass density $\rho_b=m_bc^2n_b+3n_bk_{\text{B}}T_b/2$ that is sum of rest mass and kinetic energy densities provided $k_{\text{B}}T_b/(m_bc^2)\ll 1$. As far as the nonrelativistic approximation is valid ($k_{\text{B}}T_b/(m_bc^2)\ll 1$), the equation of state for baryons reduces to $p_b\approx 0$ and $\rho_b\approx m_bc^2n_b$ where $n_b$ is the number density and $m_b$ is the rest mass. (If baryons have different masses, we sum over all baryons). Here again we have two pressureless components, the DM and baryons, with a total relative density $\Om_{0}=0.279$ at the present age.

      To this application correspond four attractors: $D$ if $\la^2=3$, $G$ and $J$ if $\la^2>3$, and $H$ if $\la^2\geq3$ and one saddle point $A$. In this case, the model describes the evolution of the universe starting from any point near the line through $A$ and $E$ (representing $I$), where pressureless matter dominates, and ending up on the line through $H$ and $G$ (representing $J$) as shown in Fig~\ref{Fig2}.
  \item $\ga<2/3$. We have seen that this is the case where the universe may undergo (at least) two TPA's and two TPD's. In this case the barotropic fluid, as the scalar field, has a negative pressure too. Arguing that each component with negative pressure causes a TPA to occur in the history of the universe, we may consider the barotropic fluid as another source of DE. Both sources of DE acting together can be understood as a rough approximation to a more general and elaborate source of DE.

      To this application correspond two attractors: $D$ if $\la^2=3$ or $\la^2=3\ga$, and $G$ if $\la^2>3\ga$. The attractor $G$ is rather a scaling solution of these two sources of DE. For instance, we may have an evolution from $E$ to $G$ or from $H$ (unstable in this case) to $G$ as Fig~\ref{Fig6} shows.

      However, to have a faithful description of the evolution of the universe one should introduce an ordinary matter or baryonic component $\rho_b$ (radiation may be neglected). For a pressureless matter component all that one needs is to add the extra equation
      \begin{equation}\label{ex}
        u'=3u [x^2-y^2+(\gamma -1) z^2]/2,\qquad (\ga<2/3),
      \end{equation}
      to the system~\eqref{6} to~\eqref{8} with $u=\ka\sqrt{\rho_b}/(\sqrt{3}H)$ and $x^2+y^2+z^2+u^2+w^2=1$.

\end{enumerate}

These are the known cases where the barotropic fluid has applications in the context of a \P. The case of a kination or stiff matter, which corresponds to $\ga=2$, may be relevant at early times if interactions are taken into considerations. However, some authors argue that interactions could still be neglected in this case and considered a (massless and free) kination along with a DE-scalar-field component with exponential potential~\cite{C}. The case $\ga=2$ would generalize the investigation made in~\cite{C} by including a non-interacting DM component. Specifically, this generalizes the two repellers $B_{\pm}$, which correspond to singularities in the scalar field, to the semi-circle $B_+EB_-$ of Fig.~\ref{Fig5} and generalizes the scaling solution $a(t)\propto t^{2/\la^2}$ of~\cite{C}, which becomes now stable for $\la^2\leq 3$ (table~\ref{Tab1}, the CP $D$).

\section{Concluding remarks \label{sec6}}

We have generalized the results obtained in~\cite{p1} and derived new ones. The conclusions we could reach are: (1) The scalar field dominated solution is stable for $\la^2\leq\min(3,3\ga)$ (this was stable for $\la^2<3\ga$~\cite{p1}). (2) The potential-kinetic-matter scaling solution is stable for $\ga\leq 1$ (its corresponding solution~\cite{p1} is a potential-kinetic scaling one the stability of which does not depend on $\ga$). This constituted the main solution derived in~\cite{p1}. With the inclusion of DM, this solution is no longer stable for $\ga>1$, no longer an attractor; rather, it is a saddle point (table~\ref{Tab1}, the CP $G$) and thus a transient potential-kinetic-radiation (taking $\ga=4/3$) equilibrium point. Such possibility is not offered in the 2-fluid problem. The derivation of (3) new attractors (the potential-kinetic-DM scaling solution and the potential-kinetic-matter-DM scaling solution), (4) new repellers and saddle points, and (5) solutions with one and two TPA's and one and two TPD's.

We have obtained attractor solutions where both DE and DM coexist and the late-time density is shared according to $\Om_{\phi}=3/\la^2$ and $\Om_{\phi}+\Om_{\text{DM}}=1$ in a way independent of the value of $\ga >1$. The case of a pressureless barotropic fluid ($\ga=1$) is more interesting and has a one-parameter family of attractors where all components coexist with, as before, $\Om_{\phi}=3/\la^2$ but $\Om_{\ga}+\Om_{\text{DM}}=1-(3/\la^2)$. New one-parameter families of matter-DM saddle points and kinetic-matter repellers were also derived. The ten CP's may be grouped into families as follows.

(1) Repellers. These include $B_{\pm}$ and $F$ and they are represented by the semicircle of Fig.~\ref{Fig5}, which includes $E$ if $\ga =2$. Eqs.~\eqref{14a}-~\eqref{14c} imply $\Om_{\text{DM}}'=3\Om_{\text{DM}}$ and $\Om_{\ga}'=3(2-\ga)\Om_{\ga}$. Thus, for $B_{\pm}$ both relative densities, $\Om_{\text{DM}}$ and $\Om_{\ga}$, increase at the beginning of the evolution against $\Om_{\phi}$ which starts decreasing. This applies to $F$ too with the exception that $\Om_{\ga}$ has a stationary value at the beginning of the evolution. In Fig.~\ref{Fig1} we plot three solution curves two of which come very close to $A$ then converge to $H$.

(2) Saddle points. If $\ga=1$, these include all the points on the line through $A$ and $E$ (representing $I$). For $0<\ga <2$, $E$ is a saddle point. $D$, $G$ and $H$ behave under some parameter restrictions as saddle points too, as shown in table~\ref{Tab1}. They all have different values of $q$. $\of$ remains undetermined. For $\ga=1$, as solution curves approach $I$, as shown in Fig.~\ref{Fig2}, all relative densities tend to become stationary as their derivatives vanish there by~\eqref{14a}-~\eqref{14c}. Thus, $I$ is a turning point. This is almost obvious from Fig.~\ref{Fig2} where the two curves, which start from $B_+$ and converge to two different values of $J$ (here $J$ is a one-parameter family of attractors which is a vertical line through $H$ and $G$ in the phase diagram), have their maximum values of $z$ ($\Om_{\ga}=z^2$) in the vicinity of $I$.

We have also noticed that, for $0\leq \ga < 2/3$, a TPA occurs as the state of the universe approaches the intermediate state defined by the saddle point $E$ (where both kinetic and potential energies are zero), which lasts longer than other TPA's occurring away from saddle points (where the kinetic energy has a minimum and the potential energy has a maximum). To our knowledge, such a conclusion was never discussed in other \P s with interactions.

(3) Attractors. $J$ and $G$ form a set of attractors for $\ga\leq 1$ and $\la^2 >3\ga$. $J$ is a one-parameter family of new attractors represented by a vertical line in the phase diagram which extends from the point $H=\sqrt{3/2}/\la,\,\sqrt{3/2}/\la,\,0)$ to the point $G=(\sqrt{3/2}/\la,\,\sqrt{3/2}/\la,\,\sqrt{1-(3/\la^2)})$. For $\ga>1$, we have the potential-kinetic-DM scaling solution, $H$, which is a new attractor where the end-behavior of the universe's evolution does not depend on the barotropic fluid equation of state. To our knowledge, this point was never discussed in other \P s with interactions. $G$, the potential-kinetic-matter scaling solution, is stable for $\ga \leq 1$ but the universe approaching this late-time state undergoes a decelerated expansion, as it should be, only if $(1\geq)\ga >2/3$.

It is straightforward to see that the CP's correspond to power law solutions for the scale factor $a(t)$, as is the case with the CP's of the 2-fluid problem~\cite{p1,p2}. If ($x_c\neq 0$ and $z_c\neq 0$), we obtain from~\eqref{10a}  $a(t)\propto t^m$ with $m=2/(6x_c^2+3\ga z_c^2)$~\cite{p2}.

For the case $\ga =1$, it is interesting to give a qualitative description of the solution curves which lie on the ellipsoid~\eqref{13}. For $\ga =1$, Eq.~\eqref{11} implies $(r^2)'=3(r^2-1)(x^2-y^2)$. Thus, curves with higher kinetic energy densities ($x^2>y^2$) move upward on the ellipsoid, in the direction of decreasing $r$ i.e. decreasing $\Om_{\phi}$ and increasing $\Om_{\ga}=L^2\Om_{\text{DM}}$, and those with lower kinetic energy densities ($x^2<y^2$) move downward in the direction of increasing $\Om_{\phi}$ and decreasing $\Om_{\ga}=L^2\Om_{\text{DM}}$. The only critical point that lies on the ellipsoid is the point $J_{\text{ellipsoid}}=(3/2/\la^2,\,3/2/\la^2,\,\ell \sqrt{1-3/\la^2})$, which also lies on the line $J$ through the points $H$ and $G$, with $w_c=\sqrt{1-\ell^2}\sqrt{1-3/\la^2}$ where $0<\ell<1$ is still a free parameter. $J_{\text{ellipsoid}}$ lies on the segment of the ellipse that joins the points $(1/\sqrt{2},\,1/\sqrt{2},\,0)$ and $(0,\,0,\,\ell)$. All solution curves end up, directly or spiraling, at $J_{\text{ellipsoid}}$. Thus, for $\la^2>24/7$, since $J$ is a stable spiral, the three relative densities, ($\Om_{\phi},\,\Om_{\ga},\,\Om_{\text{DM}}$) undergo oscillations around their average values, ($3/\la^2,\,\ell^2(1-3/\la^2),\,(1-\ell^2)(1-3/\la^2)$), respectively.

This cosmological model of three fluids, consisting of a barotropic fluid with an equation-of-state parameter $\gamma-1$, a pressureless DM fluid, plus a scalar field $\phi$ coupled to exponential potential $V=V_0\exp{(-\kappa\lambda\phi)}$, offers more possibilities for alleviating the coincidence problem: The late-time state is a decelerated expansion if $\ga >2/3$, $\of\leq 0$, and the late-time relative densities are constant (but depend on $\la$) or arbitrary with their values determined through observations only. The model fits well the three physical scenarios: $\ga=4/3$, $\ga=1$ and $\ga<2/3$ as discussed in Sect.~\ref{sec5}.


\end{document}